\documentclass[epj]{svjour}
\usepackage{latexsym,epsfig}
\usepackage{graphics,graphicx}

\newcommand \tie {{\it i.e.}}

\newcommand \kd  {\delta}
\newcommand \ra  {\rightarrow}

\newcommand \fx {{\bf x}}

\newcommand \g {\gamma}

\newcommand \x {\cdot}

\newcommand \A {\alpha}

\newcommand \lc {\langle}
\newcommand \rc {\rangle}

\newcommand \D {\Delta}

\newcommand \nt {\noindent}

\newcommand \bvec{\left( \begin{array}{c} }
\newcommand \evec{\end{array} \right)}

\newcommand \bea{\begin{eqnarray} }
\newcommand \eea{\end{eqnarray} }
\newcommand \nn {\nonumber}
\newcommand {\be} {\begin{equation}}
\newcommand {\ee} {\end{equation}}
\newcommand {\epem} {$e^+ e^-$}

\begin{document}
\title{Dihadron fragmentation: in vacuum and in matter}
\author{A. Majumder\inst{1}} 
%
%
\institute{Nuclear Science Division,
Lawrence Berkeley National Laboratory,
1 Cyclotron road, Berkeley, CA 94720}
\date{Received: date / Revised version: date}
%
\abstract{Two particle correlations within a single jet produced 
in deeply inelastic scattering (DIS) off a large nucleus as 
well as in heavy-ion collisions are explored. This 
is performed within the framework of the 
medium modified dihadron fragmentation functions.
The modification occurs due to
gluon bremsstrahlung induced by multiple scattering. 
The modified fragmentation
functions for dihadrons are found to follow closely that of single
hadrons leading to a weak nuclear suppression of their ratios as
measured by HERMES in DIS experiments. Meanwhile, a moderate
medium enhancement of the near-side correlation of two high $p_T$
hadrons is found in central heavy-ion collisions, partially due to
trigger bias caused by the competition between parton energy loss
and the initial Cronin effect.
} 
\maketitle
%

%

\section{Introduction}

The modification of the properties of jets as they pass through dense 
matter has emerged as a new diagnostic tool in the study of the partonic 
structure of such an environment \cite{Gyulassy:2003mc}. Such modification goes beyond a 
mere suppression of the directed energy leading to a suppression of the 
multiplicity \cite{Gyulassy:2003mc,guowang} 
and could in principle be extended to include the modification 
of many particle observables. Theoretically, such observables may be 
computed through a study of $n$-hadron fragmentation functions. Such 
functions may be defined as the $n$-hadron expectation values of partonic 
operators. These may be factorized from the hard collision and their 
evolution with energy scale and modification in matter may be 
systematically studied in perturbative Quantum Chromodynamics (pQCD) \cite{guowang,amxnw}. 
This is quite similar to the usual single inclusive
hadron fragmentation functions \cite{col89}.


Experimentally, such objects may be estimated through the measurement of 
semi-inclusive distributions of hadrons emanating from such reactions.
Along with the multiplicity distribution at a given momentum or momentum
fraction, which corresponds to the single fragmentation function, a variety 
of triggered distributions may also be measured. In such experiments, 
the measurement of one or more associated hadrons is only undertaken 
in the event of the appearance of a trigger hadron with a certain characteristic
momentum or momentum fraction. Such measurements have recently been performed in 
DIS off cold nuclear matter by the HERMES 
experiment at DESY \cite{dinezza04}.
Measurements in hot and dense nuclear matter have been performed at the 
Relativistic Heavy-Ion Collider (RHIC) by the STAR and PHENIX detectors \cite{star2,phenix2}.  
Due to issues related to background subtraction and requisite statistics, 
such measurements were restricted to two hadron correlations. 
While the two-hadron correlation
is found to be slightly suppressed in DIS off a nucleus versus
a nucleon target, it is moderately enhanced in central $Au+Au$
collisions relative to that in $p+p$. This is in sharp contrast
to the observed strong suppression of single inclusive
spectra \cite{hermes1,highpt} in both DIS and central $A+A$ collisions.
The theoretical description of this behaviour forms the topic of this 
effort. 


\section{Definitions}

Similar to single hadron fragmentation functions,
dihadron fragmentation functions of a quark can be defined as the
overlapping matrices of the quark fields and two-hadron final states,

\bea
D_q^{h_1,h_2}(z_1,z_2) &=& \frac{z^4}{4z_1z_2} 
\int \frac{d^2q_\perp}{4(2\pi)^3} 
\int \frac{d^4 p}{(2\pi)^4} \kd \left( z - \frac{p_h^+}{p^+}  \right)  \nn \\ 
&\times& {\rm Tr} \Bigg[ \frac{\g^+}{2p_h^+} 
\int d^4 x e^{i p \x \fx} \sum_{S - 2} \nn \\
& & \hspace{-1.0in} \times 
\lc 0 | \psi_q (x) | p_1, p_2, S-2 \rc  
\lc p_1, p_2, S-2 | \bar{\psi}_q (0) | 0 \rc \Bigg].
\eea  
and can be factorized from the hard processes \cite{amxnw},
where $p_h^+ \equiv p_1^+ + p_2^+$ is the total momentum
of the two hadrons with flavors $h_1$ and $h_2$,
$z \equiv z_1 + z_2$ is the corresponding total momentum fraction
and $q_{\perp}=p_{1 \perp}-p_{2 \perp}$ is the relative transverse
momentum.

They also satisfy the Dokshitzer-Gribov-Lipatov-Altarelli-Parisi (DGLAP)
evolution equations as have been derived in Ref.~\cite{amxnw}.
One unique feature of the DGLAP equations for dihadron
fragmentation functions is the contribution from independent
fragmentation of two partons after the parton splitting,
which involves the product of two single fragmentation functions.
These DGLAP equations for dihadron fragmentation functions
have been solved numerically \cite{Majumder:2004br} and the $Q^2$ evolution
of the dihadron fragmentation functions agrees very well with results
from JETSET \cite{jetset} Monte Carlo simulations of
$e^+ + e^- \rightarrow h_1+h_2+ X$ processes. Although both
the single and dihadron fragmentation functions evolve rapidly
with $Q^2$, their ratio has a very weak $Q^2$ dependence.

As in the case for the single fragmentation functions, 
dihadron fragmentation functions 
may not be predicted entirely within QCD, but need to be measured 
for all momentum fractions at a given energy scale. Since there 
are no experimental data available for dihadron
fragmentation functions, JETSET Monte Carlo results will be used as
the initial condition for the vacuum dihadron fragmentation
functions in this study. For single hadron fragmentation functions,
the BKK parameterization \cite{bin95} will be used; this also agrees
well with JETSET results.

Besides predicting the factorization and evolution of appropriately 
defined parton matrix elements, pQCD also predicts a universality of these 
matrix elements. The fragmentation functions defined in \epem collisions are 
identical to that in Deep-Inelastic Scattering (DIS) and in hadron-hadron 
or nucleus-nucleus collisions. The hard scattering cross section ($\sigma_0$) 
is 
however different in each case.

\section{Modification in a cold medium}

Applying factorization to dihadron production in single jet events in
DIS off a nucleus, $e(L_1)+A(p)\rightarrow e(L_2)+h_1(p_1)+h_2(p_2) +X$,
one can obtain the dihadron semi-inclusive cross section as,
\begin{equation}
E_{L_2}\frac{d\sigma^{h_1h_2}_{\rm DIS}}{d^3L_2 dz_1dz_2}
=\frac{\alpha^2}{2\pi s}\frac{1}{Q^4}L_{\mu\nu}\frac{dW^{\mu\nu}}{dz_1dz_2},
\end{equation}
in terms of the semi-inclusive tensor at leading twist,
\begin{eqnarray}
\frac{d W^{\mu \nu } }{d z_1 d z_2} &=&\sum_q \int d x  f^A_q(x,Q^2)
H^{\mu \nu}(x,p,q) \nonumber \\
&\times& D_q^{h_1,h_2}(z_1,z_2,Q^2).
\end{eqnarray}
In the above, $D_q^{h_1,h_2}(z_1,z_2)$ is the dihadron fragmentation
function,
$L_{\mu\nu}=(1/2){\rm Tr}(\not\!\!L_1\gamma_\mu\not\!\!L_2\gamma_\nu$),
the factor $H^{\mu \nu}$ represents the hard part of quark
scattering with a virtual photon which carries a four-momentum
$q=[-Q^2/2q^-,q^-,\vec{0}_\perp]$ and $f^A_q(x,Q^2)$ is the quark
distribution in the nucleus which has a total momentum
$A[p^+,0,\vec{0}_\perp]$. The hadron momentum fractions,
$z_1=p_1^-/q^-$ and $z_2=p_2^-/q^-$, are defined with respect to
the initial momentum $q^-$ of the fragmenting quark.

At next-to-leading twist, the dihadron semi-inclusive tensor
receives contributions from multiple scattering of the struck
quark off soft gluons inside the nucleus with induced gluon
radiation. One can reorganize the total contribution
(leading and next-to-leading twist) into a product of effective
quark distribution in a nucleus, the hard part of photon-quark
scattering $H^{\mu \nu}$ and a modified dihadron fragmentation
function $\tilde{D}_q^{h_1,h_2} (z_1,z_2)$. The calculation of
the modified dihadron fragmentation function at the
next-to-leading twist in a nucleus proceeds \cite{maj04f}
similarly as that for the modified single hadron fragmentation
functions \cite{guowang} and yields,
\bea
\tilde{D}_q^{h_1,h_2} (z_1,z_2) &=& D_q^{h_1,h_2}(z_1,z_2) +
\int_0^{Q^2} \frac{dl_{\perp}^2}{l_{\perp}^2} \frac{\A_s}{2\pi}
 \nn \\
& &\hspace{-1in} \times\left[ \int_{z_1+z_2}^1 \frac{dy}{y^2}
\left\{ \D P_{q\ra q g} (y,x_B,x_L,l_\perp^2)
D_q^{h_1,h_2} \left(\frac{z_1}{y},\frac{z_2}{y} \right)\right.\right. \nn \\
& &\hspace{-0.8in}+\left. \D P_{q\ra g q} (y,x_B,x_L,l_\perp^2)
D_g^{h_1,h_2} \left(\frac{z_1}{y},\frac{z_2}{y} \right) \right\} \nn \\
& & \hspace{-0.8in}+\int_{z_1}^{1-z_2} \frac{dy}{y(1-y)}
\D \hat{P}_{q\ra q g} (y,x_B,x_L,l_\perp^2) \nn \\
& &\hspace{-0.8in}\times  \left. D_q^{h_1}
\left(\frac{z_1}{y})\right) D_g^{h_2}\left(\frac{z_2}{1-y} \right)
+ (h_1 \ra h_2) \right]\, .
\label{eq-dihdr-mod}
\eea
In the above, $x_B= -Q^2/2p^+q^-$, $x_L = l_\perp^2/2p^+q^-y(1-y)$,
$l_\perp$ is the transverse momentum of the radiated gluon,
$\D P_{q\ra qg}$ and $\D P_{q\ra g q}$
are the modified splitting functions whose forms are identical
to that in the modified single hadron fragmentation functions \cite{guowang}.
The switch $(h_1 \ra h_2)$ is only meant for the last term, which
represents independent fragmentation of the quark and gluon after
the induced bremsstrahlung. The corresponding modified splitting
function,
\bea
\D \hat{P}_{q\ra gq} = \frac{1+y^2}{1-y}
\frac{C_A 2\pi \A_s T^A_{qg} (x_B,x_L)}{(l_\perp^2+\lc k_\perp^2\rc)
N_c f_q^A(x_B,Q^2)},
\eea
is similar to $\D P_{q\ra qg}$ but does not contain
contributions from virtual corrections.
In the above, $C_A=3$, $N_c=3$, and $\lc k_\perp^2\rc$ is the
average intrinsic parton transverse momentum inside the nucleus.

Note that both modified splitting functions depend on the
quark-gluon correlation function  in the nucleus $T^A_{qg}$, which 
also determines the modification of the single hadron fragmentation
functions \cite{maj04f}. For a Gaussian nuclear distribution, it
can be estimated as \cite{guowang}
\begin{equation}
T^A_{qg}(x_B,x_L) = \tilde{C}(Q^2) m_N R_A f_q^A(x_B) (1 - e^{-x_L^2/x_A^2}),
\label{eq-tqg}
\end{equation}
where, $x_A=1/m_NR_A$, $m_N$ is the nucleon mass and $R_A = 1.12
A^{1/3}$ is the nuclear radius. The overall constant
$\tilde{C}\propto x_\perp G(x_\perp)$ [$x_\perp=\lc
k_\perp^2\rc/2p^+q^-y(1-y)$] is the only parameter in the modified
dihadron fragmentation function which might depend on the
kinematics of the DIS process but is identical to the parameter in
the modified single fragmentation functions. In the
phenomenological study of the single hadron fragmentation
functions in DIS off nuclei, $\tilde{C}=0.006$ GeV$^2$ is
determined within the kinematics of the HERMES experiment
\cite{hermes1}. The predicted dependence of nuclear modification
of the single hadron fragmentation function on the momentum
fraction $z$, initial quark energy $\nu=q^-$ and the nuclear size
$R_A$ agrees very well with the HERMES experimental data
\cite{EW1}. 

\begin{figure}[htbp]
\hspace{0.5cm}
\epsfig{file=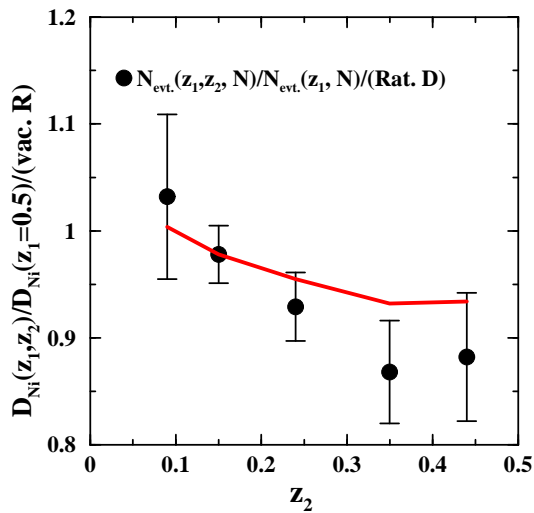,scale=0.4}
    \caption{ Results of the medium modification of
    the associated hadron distribution, in a cold nuclear medium
    versus its momentum fraction. The momentum fraction of the leading hadron
    $z_1$ is integrated over
    all allowed values above 0.5.
    The ratio of this quantity in Nitrogen (N) with that in
    deuterium (D) are the HERMES data points. }
    \label{fig1}
\end{figure}

With no additional parameters in Eq.~(\ref{eq-dihdr-mod}),
one can predict the nuclear modification of dihadron fragmentation
functions within the same kinematics. Since dihadron fragmentation
functions are connected to single hadron fragmentation functions via
sum rules \cite{Majumder:2004br}, it is more illustrative to study the
modification of the conditional distribution for the second rank hadrons,
\begin{eqnarray}
R_2(z_1,z_2)\equiv D_q^{h_1,h_2}(z_1,z_2)/D_q^{h_1}(z_1),
\label{eq-corr}
\end{eqnarray}
where $z_1$ and $z_2<z_1$ are the momentum fractions of the
triggered (leading) and associated (secondary) hadrons, respectively.
Shown in Fig.~\ref{fig1} is the predicted ratio of the associated
hadron distribution in DIS off a nitrogen ($A=14$) target to
that off a proton ($A=1$), as compared to the HERMES experimental
data \cite{dinezza04}.

The modification of the dihadron fragmentation function in a nucleus with atomic number $A$ is estimated 
experimentally by the HERMES experiment \cite{dinezza04,hermes1} by measuring 
the number of events with at least two hadrons with momentum fractions $z_1,z_2$ ($N^2_A(z_1,z_2)$), 
and the number of events with at least one hadron with momentum fraction $z$ ($N_A(z)$). In an 
effort to eradicate systematic errors, a double ratio is presented (using similar measurements 
off deuterium, \tie, $A=2$), 

\bea
\mathcal{R}_2(z_2) = \frac{\frac{ \sum_{z_1>0.5} N^2_A(z_1,z_2)  }{ \sum_{z>0.5} N_A (z) }}
{\frac{\sum_{z_1>0.5} N^2_2(z_1,z_2)}{ \sum_{z>0.5} N_2 (z) }}.
\eea
\nt
This double ratio for different $z_2$ is plotted in Fig.~\ref{fig1} as the circular points..

In comparison with the experiment we present the associated hadron 
distribution: the ratio of the 
modified dihadron to the single hadron fragmentation function in a nucleus 
normalized by the same ratio in the vacuum.
As in the experiment, we integrate over $z_1>0.5$ in both the single
and dihadron fragmentation functions. The agreement between the prediction
and the data is remarkable given that no free parameters are used.
The suppression of the associated hadron distribution 
at large $z_2$ due to multiple scattering and induced gluon
bremsstrahlung in a nucleus is quite small compared to the
suppression of the single fragmentation functions \cite{hermes1,EW1}.
Since $\mathcal{R}_2$ is the ratio of double and single hadron
fragmentation functions, the effect of induced gluon radiation or
quark energy loss is mainly borne by the single spectra of the leading
hadrons. This is similar to the evolution of the dihadron fragmentation
 function with
momentum scale in the vacuum \cite{amxnw,Majumder:2004br}. At small values
of $z_2$, the modified dihadron fragmentation function rises above
its vacuum counterpart more than the single fragmentation
functions. This is due to the new contribution where each of the
detected hadrons emanates from the independent fragmentation of
the quark and the radiated gluon.


\section{Modification in a Hot medium}

In high-energy heavy-ion (or $p+p$ and $p+A$) collisions, jets are
always produced in back-to-back pairs. Correlations of two
high-$p_T$ hadrons in azimuthal angle generally have two Gaussian
peaks \cite{star2,phenix2}. Relative to the triggered hadron,
away-side hadrons come from the fragmentation of the away-side jet
and are related to single hadron fragmentation functions. On the
other hand, near-side hadrons come from the fragmentation of the
same jet as the triggered hadron and therefore
are related to dihadron fragmentation functions.

To extend the study of intrajet dihadron correlations to
heavy-ion collisions, one can simply assume $\lc k_\perp^2 \rc
\simeq \mu^2$ (the Debye screening mass) and that $\tilde{C}$ in
Eq.~(\ref{eq-tqg}) is proportional to the local gluon density of
the produced dense matter. In addition, one also has to include
the effect of thermal gluon absorption \cite{EW2} such that the
effective energy dependence of the energy loss will be different.
Such a procedure was applied to the
study of the modification of the single hadron fragmentation functions and
it successfully described the quenching of single inclusive hadron
spectra, their azimuthal anisotropy and suppression of away-side
high $p_T$ hadron correlations \cite{xnwang03}. The change of
the near-side correlation due to the
modification of dihadron fragmentation functions in heavy-ion
collisions can be similarly calculated.

The near-side correlation with background
subtracted and integrated over the azimuthal angle \cite{fqwang}
can be related to the associated hadron distribution or
the ratio of dihadron to the single hadron fragmentation functions as in
Eq.~(\ref{eq-corr}). However, one has to average over the
initial jet energy weighted with the corresponding production
cross sections. For a given value of $p_T^{\rm trig}$ of the
triggered hadron, one can calculate the average initial jet
energy $\lc E_T \rc$. Because of trigger bias and
parton energy loss, $\lc E_T \rc$ in heavy-ion collisions
is generally larger than that in $p+p$ collisions for a fixed
$p_T^{\rm trig}$. 
In these proceedings, we simply use the
calculated $\lc E_T \rc$ and its centrality dependence
for a given $p_T^{\rm trig}$ from Ref.~\cite{Wang:2003aw}
which in turn determines
the average value of $z_1=p_T^{\rm trig}/\lc E_T \rc$
and $z_2=p_T^{\rm assoc}/\lc E_T \rc$. In the same calculation,
one can also obtain the average parton energy loss as the
difference of the average initial parton energy with and
without parton energy loss,
$\lc \D E \rc=\lc E_T \rc^{\rm loss}-\lc E_T \rc^{\rm no-loss}$,
which is completely determined from the suppression of single
inclusive hadron spectra and away-side correlation. Since the
high $p_T^{\rm trig}$ biases the jet production position
towards the surface of the overlapped region, $\lc \D E \rc$
associated with a triggered hadron is always smaller than the
average energy loss of both the away-side jet and jets
in non-triggered events.

\begin{figure}[htbp]
\begin{center}
\epsfig{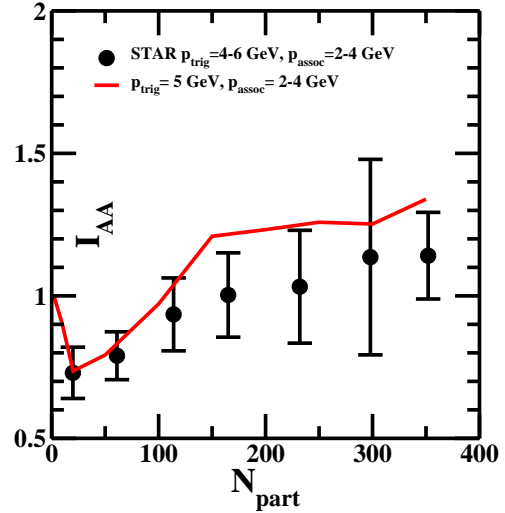}
\caption{ Calculated medium modification of associated hadron
  distribution from jet fragmentation in $Au+Au$ collisions
  at $\sqrt{s}=200$ GeV. \protect\cite{star2}. }
\end{center}  
    \label{fig2}
\end{figure}

Using $\lc \D z \rc  = \lc \D E \rc/\lc E_T \rc$, we first
determine the parameter $\tilde{C}$ 
and in turn calculate both the modified single and dihadron
distributions. The ratio of such associated hadron distributions
in $Au+Au$ versus $p+p$ collisions, referred to as $I_{AA}$
\cite{star2}, is plotted as the solid line in Fig.~\ref{fig2}
together with the STAR data \cite{star2}, as a function of the
number of participant nucleons.  
In central $Au+Au$ collisions, triggering on a high $p_T$ hadron 
biases toward a larger initial jet energy and therefore smaller 
$z_1$ and $z_2$. This leads to an enhancement in $I_{AA}$ due to
the shape of the dihadron fragmentation functions \cite{amxnw}.
This is in contrast to the slight nuclear
suppression of the associated hadron distribution at large $z_2$
in DIS off a nucleus (Fig.~\ref{fig1}).
The enhancement increases with $N_{part}$ because of increased 
initial gluon density and system size of the
produced dense matter which leads to an increased
total energy loss. In the most peripheral collisions, the effect of
smaller energy loss is countered by the Cronin effect due
to initial state multiple scattering that biases toward smaller
$\lc E_T \rc$ relative to $p+p$ collisions. As a result, the
associated hadron distribution is slightly suppressed. 

\section{Conclusions}

In summary, we have studied the medium modification of dihadron
fragmentation functions in both hot and nuclear matter due to
multiple parton scattering and induced gluon radiation. The modification
is found to follow closely that of single hadron fragmentation
functions so that the associated hadron distributions or the ratios
of dihadron and single fragmentation functions are only slightly
suppressed in DIS off nuclei but enhanced in central heavy-ion
collisions due to trigger bias. With no extra parameters, our
calculations agree very well with experimental data.
   
This work was supported by the U.S. Department of Energy under 
Contract No. DE-AC03-76SF00098, NSFC under project Nos. 10475031 
and 10135030 and NSERC of Canada.

\end{document}